\documentclass[prl,twocolumn]{revtex4}
\usepackage{amsmath} 
\usepackage{graphicx}
\usepackage{color}

\begin{document}

\title{Stochastic thermodynamics for self-propelled particles}
\author{Grzegorz Szamel}
\affiliation{Department of Chemistry, 
Colorado State University, Fort Collins, CO 80523}

\date{\today}

\begin{abstract}
We propose a generalization of stochastic thermodynamics to systems of active
particles, which move under the combined influence of stochastic internal 
self-propulsions (activity) and a heat bath. The main idea 
is to consider joint trajectories of
particles' positions and self-propulsions. It is then possible to exploit formal
similarity of an active system and a system consisting of two subsystems interacting
with different heat reservoirs and coupled by a non-symmetric interaction. The resulting
thermodynamic description closely follows the standard stochastic thermodynamics. In 
particular, total entropy production, $\Delta s_\text{tot}$, can be decomposed into
housekeeping, $\Delta s_\text{hk}$, and excess, $\Delta s_\text{ex}$, parts.
Both $\Delta s_\text{tot}$ and $\Delta s_\text{hk}$ satisfy fluctuation theorems.
The average rate of the steady-state housekeeping entropy production can be related
to the violation of the fluctuation-dissipation theorem via a Harada-Sasa 
relation. The excess entropy production enters into a Hatano-Sasa-like relation, which
leads to a generalized Clausius inequality involving the 
change of the system's entropy and the excess entropy production. Interestingly, 
although the evolution of particles' self-propulsions is free and uncoupled from that
of their positions, non-trivial steady-state correlations between these variables
lead to the non-zero excess dissipation in the reservoir coupled to the 
self-propulsions.
\end{abstract} 

\maketitle
\textit{Introduction. --}
Active matter systems \cite{Ramaswamyrev1,Catesrev,Marchettirev1,Bechingerrev,
Ramaswamyrev2,Marchettirev2} consist of particles that move on their own accord by
consuming energy from their environment. Since these systems are
out of equilibrium, they cannot be described using standard thermodynamics.
Recently, there have been several attempts 
\cite{Fodor,Mandal,Marconi2017,Speck2016,Speck2018,Shankar,Dabelow2019,Caprini2019,
Chakrabarti,Chaudhuri} to 
extend the formalism of stochastic thermodynamics to describe active matter. 

Stochastic thermodynamics \cite{SeifertRPP} is the most successful framework
for the description of an admittedly limited class of non-equilibrium systems. In
its standard form, stochastic thermodynamics  
deals with systems which are in contact with a heat reservoir
characterized by a constant temperature and are displaced out of equilibrium 
by external forces (in contrast to active systems that evolve under the
influence of internal self-propulsions). 
It combines stochastic energetics \cite{Sekimoto1998}, 
which generalizes notions of work and heat to the level of 
systems' trajectories, with the notion of stochastic entropy \cite{Seifert2005}.
It allows one to derive a number of results characterizing out-of-equilibrium
states and processes. We shall mention here fluctuation theorems 
for entropy production \cite{GallavottiCohen,Kurchan,LebowitzSpohn} which shed light
on the probability of rare, 2nd law violating fluctuations, the Jarzynski 
relation \cite{Jarzynski1997} which expresses the free energy difference between two 
equilibrium states in terms of the average 
of the exponential of the work performed in a non-equilibrium 
process between the same two states and a generalization of the Jarzynski
relation to transitions between two stationary non-equilibrium states \cite{HatanoSasa}. 
Several results obtained in the framework of stochastic thermodynamics were 
experimentally verified, see, \textit{e.g.},
Refs. \cite{Trepagnier,Tietz,Speck2007,Ciliberto2017}.

Two types of a generalization of stochastic thermodynamics to active matter systems
have been proposed. Fodor \textit{et al.} \cite{Fodor} and Mandal \textit{et al.} 
\cite{Mandal} considered model \emph{athermal} active matter systems and derived two 
different expressions for the entropy production. The common feature of 
these two studies is that they map athermal systems onto equilibrium 
systems with non-conservative interactions in which the effective temperature (which 
characterizes the strength of the self-propulsions) plays the role of  
the temperature of the medium. It is not clear how to generalize these studies to active 
systems that are influenced by both self-propulsions and thermal noise \cite{heatbath}.  
In contrast, Speck  \cite{Speck2016,Speck2018}, Shankar and Marchetti \cite{Shankar}, 
and Dabelow \textit{et al.} \cite{Dabelow2019} followed the standard 
stochastic thermodynamics more closely and 
considered active matter systems in contact with a heat bath.

Here we follow the spirit of Ref. \cite{Shankar} 
and consider joint trajectories of particles' positions 
and self-propulsions \cite{Pietzonka}, for an active system in contact with
a heat bath \cite{commentnew}. This approach allows us 
to follow the standard stochastic thermodynamics framework and to generalize 
a number of its results to active matter systems. In particular, 
we divide the entropy production into the housekeeping part, which originates from
the non-equilibrium character of active matter, and the excess part, we derive a
fluctuation theorem for the housekeeping entropy production, 
and we relate the steady-state entropy
production to the experimentally measurable violation of a fluctuation-dissipation 
relation. We also obtain a generalized Clausius inequality which
extends the 2nd law of thermodynamics to active matter systems.

\textit{Model: Active Brownian particle in an external potential. --} 
To illustrate our approach we use a minimal model system consisting of a single
active Brownian particle (ABP) \cite{tenHagen} under the influence of an external force, 
in two spatial dimensions. The equations of motion read
\begin{align}\label{eom1}
\gamma_t\dot{\mathbf{r}} &=  \mathbf{F}_\lambda 
+ \gamma_t v_0 \mathbf{e} + \boldsymbol{\zeta}
& \left<\boldsymbol{\zeta}(t)\boldsymbol{\zeta}(t')\right> = 
2 \boldsymbol{I} \gamma_t T \delta(t-t'),
\\ \label{eom2}
\gamma_r\dot{\varphi} &= \eta & 
\left<\eta(t)\eta(t')\right> = 2 \gamma_r T \delta(t-t'),
\end{align}
where $\gamma_t$ and $\gamma_r$ are the translational and rotational friction
coefficients \cite{friction}, respectively, and 
$\mathbf{e} \!\equiv \!(\cos(\varphi),\sin(\varphi))$ is the orientation vector.
Next, $\mathbf{F}_\lambda$ 
is an external force, which depends on a control parameter $\lambda$,  
and may include a non-conservative component, 
and $v_0$ is the self-propulsion speed. Finally, $\boldsymbol{\zeta}$ and
$\eta$ are Gaussian white noises representing thermal fluctuations. 
We use the system of units such that the Boltzmann constant $k_B=1$. 

The model system defined through Eqs. (\ref{eom1}-\ref{eom2}) is mathematically 
equivalent to a system consisting of two subsystems, with each subsystem 
connected to its heat reservoir, and with the subsystems 
coupled through a non-symmetric, \textit{i.e.} violating Newton's 3rd law, 
interaction. In the present case of a single 
ABP the reservoirs are at the same temperature. If we were to consider an 
active Ornstein-Uhlenbeck particle (AOUP) \cite{Szamel2014,Maggi,Fodor}, 
the temperatures of the two reservoirs would be unrelated. 
Interestingly, the temperature of the reservoir
coupled to the self-propulsions, Eq. (\ref{eom2}), is different from the so-called
effective temperature of the self-propulsions \cite{Shankar}, 
$T_a=v_0^2\gamma_t\gamma_r/(2T)$. 

Systems consisting of subsystems 
coupled to different heat reservoirs have been extensively studied, both 
theoretically \cite{Jarzynski2004,VandenBroeck2004,Visco2006,Crisanti2012,Fogedby2011,
Fogedby2012,Fogedby2014}
and experimentally \cite{Ciliberto2017,CilibertoPRL3013,CilibertoJSM2013}. 
We note that in the 
present case, with a non-symmetric interaction between the two subsystems,
even if the temperatures of the reservoirs are the same, the combined system is out 
of equilibrium.

\textit{Stochastic entropy production. --}
To define a stochastic entropy we follow Seifert \cite{Seifert2005}. First,
we consider the Fokker-Planck equation for the joint probability density
for the particle's position and self-propulsion, which corresponds to 
Eqs. (\ref{eom1}-\ref{eom2}),
\begin{eqnarray}\label{FP1}
\partial_t p(\mathbf{r},\varphi;t) = 
-\partial_\mathbf{r}\cdot\mathbf{j}_t(\mathbf{r},\varphi;t)
-\partial_\varphi j_r(\mathbf{r},\varphi;t),
\end{eqnarray}
where current densities in the position and orientation spaces read
\begin{eqnarray}\label{currentr}
\mathbf{j}_t(\mathbf{r},\varphi;t) &=&
\gamma_t^{-1}\left(\mathbf{F}_\lambda+\gamma_t v_0\mathbf{e} 
- T\partial_\mathbf{r}\right)p(\mathbf{r},\varphi;t),
\\ \label{currentp}
j_r(\mathbf{r},\varphi;t)&=&
- \left(T/\gamma_r\right) \partial_\varphi p(\mathbf{r},\varphi;t).
\end{eqnarray}
We emphasize that due to the non-symmetric interaction between $\mathbf{r}$ and $\varphi$
sectors, in a steady state currents (\ref{currentr}-\ref{currentp}) do not 
vanish \cite{Grosberg}, even if $\mathbf{F}_\lambda$ is conservative. 

Next, we define the trajectory-level entropy for the system (\textit{i.e} the particle), 
using the solution of the Fokker-Planck equation (\ref{FP1}) for a time-dependent control
parameter $\lambda(t)$, evaluated along 
the stochastic trajectory of the particle's position and self-propulsion,
\begin{eqnarray}\label{sp1}
s(t) = -\ln p(\mathbf{r}(t),\varphi(t);t).
\end{eqnarray}
The rate of change of the systems entropy reads
\begin{eqnarray}\label{speom1}
\dot{s}(t) &=& -\frac{\partial_t p(\mathbf{r},\varphi;t)}{p(\mathbf{r},\varphi;t)}
-\frac{\partial_\mathbf{r} p(\mathbf{r},\varphi;t)}{p(\mathbf{r},\varphi;t)}
\cdot\dot{\mathbf{r}}
-\frac{\partial_\varphi p(\mathbf{r},\varphi;t)}{p(\mathbf{r},\varphi;t)}\dot{\varphi}
\nonumber \\ &=& 
-\frac{\partial_t p(\mathbf{r},\varphi;t)}{p(\mathbf{r},\varphi;t)}
+ \frac{\mathbf{j}_t(\mathbf{r},\varphi;t)}{(T/\gamma_t) p(\mathbf{r},\varphi;t)}
\cdot\dot{\mathbf{r}}
\nonumber \\ && 
- \frac{\mathbf{F}+\gamma_t v_0 \mathbf{e}}{T} \cdot\dot{\mathbf{r}}
+ \frac{j_r(\mathbf{r},\varphi;t)}{(T/\gamma_r) p(\mathbf{r},\varphi;t)}\dot{\varphi}.
\end{eqnarray}
The first term in the last line can be interpreted as the entropy production in the
medium,
\begin{eqnarray}\label{sm1}
\dot{s}_\text{m}(t) = T^{-1} \left(\mathbf{F}+\gamma_t v_0 \mathbf{e}\right)
\cdot\dot{\mathbf{r}}.
\end{eqnarray}
We will see in the remainder of this Rapid Communication 
that identification (\ref{sm1}) leads to 
a consistent framework of active stochastic thermodynamics. Using Eqs. 
(\ref{speom1}-\ref{sm1}) we can express the total stochastic entropy production in terms 
of local velocities evaluated along the trajectory,
\begin{eqnarray}\label{stot1}
\dot{s}_\text{tot}(t) = 
-\frac{\partial_t p(\mathbf{r},\varphi;t)}{p(\mathbf{r},\varphi;t)}
+ \frac{1}{T}\left(\gamma_t \mathbf{v}_{t}\cdot\dot{\mathbf{r}}
+ \gamma_r v_{r} \dot{\varphi}\right),
\end{eqnarray}
where $\mathbf{v}_{t}(\mathbf{r},\varphi;t) =
\mathbf{j}_{t}(\mathbf{r},\varphi;t)/p(\mathbf{r},\varphi;t)$ and
$v_{r}(\mathbf{r},\varphi;t) =
j_{r}(\mathbf{r},\varphi;t)/p(\mathbf{r},\varphi;t)$.

We note that averaging $\dot{\mathbf{r}}$ and $\dot{\varphi}$  over all 
trajectories under the condition that the position and self-propulsion at time $t$ 
are equal to $\mathbf{r}$ and $\varphi$ gives the local translational 
and rotational velocity, respectively,
\begin{eqnarray}\label{velr}
\left<\dot{\mathbf{r}}|\mathbf{r},\varphi\right> &=& 
\mathbf{v}_{t}(\mathbf{r},\varphi;t) 
\\ \label{velp}
\left<\dot{\varphi}|\mathbf{r},\varphi\right> &=&
v_{r}(\mathbf{r},\varphi;t).
\end{eqnarray} 
Here and in the following $\left<\cdots\right>$ denotes averaging over the 
trajectories.

Combining Eqs. (\ref{stot1}) and (\ref{velr}-\ref{velp}) allows us 
to calculate the average total entropy production, 
$\dot{S}(t) = \left<\dot{s}_\text{tot}(t)\right>$,
\begin{eqnarray}\label{stotave}
\dot{S}(t) = 
\int d\mathbf{r}d\varphi \frac{\gamma_t \mathbf{v}_t^2(\mathbf{r},\varphi;t) + 
\gamma_r v_r^2(\mathbf{r},\varphi;t)}{T} p(\mathbf{r},\varphi;t).
\nonumber \\
\end{eqnarray}
Due to non-vanishing currents, the total entropy increases even in a steady state 
with a conservative force. 

\textit{Housekeeping entropy production and Harada-Sasa relation --} 
Oono and Paniconi \cite{Oono} introduced
the concept of a housekeeping heat, \textit{i.e.} the heat dissipated in a
non-equilibrium steady state. Here we generalize this concept into 
a housekeeping entropy production. 
Following the spirit of Hatano and Sasa 
\cite{HatanoSasa} and of Speck and Seifert \cite{SpeckSeifert} we define the 
housekeeping increase of the entropy, $\Delta s_\text{hk}$, as follows
\begin{eqnarray}\label{shk1}
\Delta s_\text{hk} = T^{-1} \int_0^t dt' 
\left(\gamma_t \mathbf{v}_{ts}\cdot\dot{\mathbf{r}}
+ \gamma_r v_{rs} \dot{\varphi}\right).
\end{eqnarray}
Here $\mathbf{v}_{ts}$ and $v_{rs}$ are steady-state local translational and rotational
velocities, respectively,
\begin{eqnarray}\label{velrs}
\mathbf{v}_{ts}(\mathbf{r},\varphi|\lambda(t)) &=& 
\mathbf{j}_{ts}(\mathbf{r},\varphi|\lambda(t))/p_s(\mathbf{r},\varphi|\lambda(t))
\\ \label{velps}
v_{rs}(\mathbf{r},\varphi|\lambda(t)) &=& 
j_{rs}(\mathbf{r},\varphi|\lambda(t))/p_s(\mathbf{r},\varphi|\lambda(t)),
\end{eqnarray} 
evaluated along the stochastic trajectory. In Eqs. (\ref{velrs}-\ref{velps})
$\mathbf{j}_{ts}(\lambda(t))$, $j_{rs}(\lambda(t))$ and $p_s(\lambda(t))$ 
are the currents and 
the probability distribution in a steady state corresponding to a fixed instantaneous 
value of the control parameter, $\lambda(t)$. 
We note that the housekeeping entropy increase originates from \textit{both}
translational and rotational degrees of freedom, in spite of the fact that the
rotational motion is free and decoupled from the translational one. Furthermore, 
as expected, in a steady state the total entropy production
(\ref{stot1}) and the housekeeping entropy production (\ref{shk1}) coincide.

Using the regularization method described by Speck and Seifert \cite{SpeckSeifert} 
one can derive an equation of motion for the joint probability distribution of 
the particle's position, orientation and housekeeping entropy increase, 
$\rho(\mathbf{r},\varphi,\Delta s_\text{hk};t)$,
\begin{widetext}
\begin{eqnarray}\label{rhoeom1}
&& \partial_t \rho(\mathbf{r},\varphi,\Delta s_\text{hk};t) =
-\gamma_t^{-1}\partial_\mathbf{r}\cdot\left[\mathbf{F}+\gamma_t v_0\mathbf{e} 
- T\partial_\mathbf{r}\right]\rho(\mathbf{r},\varphi,\Delta s_\text{hk};t)
+(T/\gamma_r)\partial_\varphi^2 \rho(\mathbf{r},\varphi,\Delta s_\text{hk};t)
\\ \nonumber && 
+ T^{-1}\left[\gamma_t \mathbf{v}_{ts}^2 + \gamma_r v_{rs}^2 \right]
\partial^2_{\Delta s_\text{hk}}\rho(\mathbf{r},\varphi,\Delta s_\text{hk};t)
+\left[2\partial_\mathbf{r}\cdot\mathbf{v}_{ts} + 
2\partial_\varphi v_{rs} 
-T^{-1} \left(\gamma_t \mathbf{v}_{ts}^2 + \gamma_r v_{rs}^2 \right)
\right]\partial_{\Delta s_\text{hk}}
\rho(\mathbf{r},\varphi,\Delta s_\text{hk};t)
\end{eqnarray}
\end{widetext} 
Equation of motion (\ref{rhoeom1}) allows us to 
show that the average of $\exp(-\Delta s_\text{hk})$ is time-independent, 
\begin{eqnarray}\label{expshkt}
\frac{d}{dt}
\int d\mathbf{r}d\varphi d\Delta s_\text{hk} e^{-\Delta s_\text{hk}}
\rho(\mathbf{r},\varphi,\Delta s_\text{hk};t) = 0, 
\end{eqnarray}
which leads \cite{SpeckSeifert} to the integral fluctuation
theorem for the housekeeping entropy production,
\begin{eqnarray}\label{shkift}
\left<\exp(-\Delta s_\text{hk})\right> = 1.
\end{eqnarray}
We note that fluctuation theorem (\ref{shkift}) is valid for any time dependence of the 
control parameter, including the time-independent order parameter, \textit{i.e}
the steady state. 

The average steady state housekeeping entropy production, which can 
be calculated from Eq. (\ref{rhoeom1}) as
\begin{eqnarray}\label{shktave}
\partial_t \left<\Delta s_\text{hk}\right>
= T^{-1}\left<\gamma_t \mathbf{v}_{ts}^2 + \gamma_r v_{rs}^2 \right>,
\end{eqnarray}
can be related to a violation of the fluctuation-response relation \textit{via}
an equality equivalent to the Harada-Sasa relation \cite{HaradaSasa}. To prove this
equality we first consider a perturbation of our system, initially in a steady state,
by a weak, constant in space, external force $\epsilon \mathbf{f}_\text{ext}(t)$.  
The change of the average translational 
velocity of the particle can be expressed in terms
of response function $\boldsymbol{R}(t)$,
\begin{eqnarray}\label{resp}
\delta \left<\mathbf{v}_t(t)\right> = \epsilon \int_{-\infty}^t \boldsymbol{R}(t-t')\cdot
\mathbf{f}_\text{ext}(t').
\end{eqnarray}
The response function (which geometrically is a second rank tensor) has an
instantaneous part and a time delayed part. The short-time limit of the 
time-delayed response can be calculated as
\begin{eqnarray}\label{respt0}
\boldsymbol{R}(0^+) = 
-\gamma_t^{-2}\int d\mathbf{r}d\varphi
\left(\mathbf{F}+\gamma_t v_0\mathbf{e}\right)\partial_\mathbf{r} p_s(\mathbf{r},\varphi)
\end{eqnarray}
In an equilibrium Brownian system the time-dependent response function 
$\boldsymbol{R}(t)$ is related to velocity autocorrelation function,
$\boldsymbol{C}(t)=\left<\mathbf{v}(t)\mathbf{v}(0)\right>$ through the
fluctuation-dissipation relation, $T\boldsymbol{R}(t) = \boldsymbol{C}(t)$. 
The velocity autocorrelation function has a part proportional to
the $\delta$ function and a time-dependent part. For our system the short-time limit 
of the latter part can be calculated as
\begin{eqnarray}\label{vacft0}
\boldsymbol{C}(0^+) &=& 
\gamma_t^{-2}\int d\mathbf{r}d\varphi
\left(\mathbf{F}+\gamma_t v_0\mathbf{e}\right)
\\ \nonumber && \times
\left(\mathbf{F}+\gamma_t v_0\mathbf{e}-2T\partial_\mathbf{r} \right)
p_s(\mathbf{r},\varphi)
\end{eqnarray}
Combining Eqs.(\ref{respt0}) and (\ref{vacft0}) and then using the equation 
for the steady-state probability distribution one can show that in a steady state
\begin{eqnarray}\label{HSR}
\dot{S}(t) = \partial_t \left<\Delta s_\text{hk}\right> = 
(\gamma_t/T) \mathrm{Tr} \left[ \boldsymbol{C}(0^+) - T \boldsymbol{R}(0^+)\right].
\end{eqnarray}
Equality (\ref{HSR}) is the Harada-Sasa relation \cite{HaradaSasa} 
written in the time domain (note that the instantaneous part of $T\boldsymbol{R}(t)$ 
cancels the $\delta$ function part of $\boldsymbol{C}(t)$).
It expresses the average steady state entropy production in terms of the
violation of the fluctuation-dissipation relation involving the dynamics
of the particle's position. Interestingly, although the analogous 
fluctuation-dissipation relation for the dynamics of the particle's orientation
is not violated (in fact, rotational dynamics is free), the expressions
for both the total and housekeeping entropy production include terms that 
originate from the rotational motion. We note that a generalized
Harada-Sasa relation was also derived within a field-theoretical description of active
matter \cite{Nardini2017}. 

\textit{Hatano-Sasa relation and a generalized Clausius inequality. --} 
Using Eq. (\ref{sm1}) we can write the entropy dissipated into the medium as
\begin{eqnarray}\label{sm2}
\Delta s_\text{m}(t) &=& T^{-1} \int_0^t dt' 
\left(\mathbf{F}(t')+\gamma_t v_0 \mathbf{e}(t')\right)
\cdot\dot{\mathbf{r}}(t') 
\\ \nonumber &=& \Delta s_\text{hk}(t) - \Delta \phi + \int_0^t dt' 
\dot{\lambda}(t')\partial_\lambda \phi(\mathbf{r},\varphi; \lambda),
\end{eqnarray}
where $\phi(t) = - \ln p_s(\mathbf{r}(t),\varphi(t);\lambda(t))$ and
$\Delta \phi = \phi(t)-\phi(0)$ \cite{phistochentr}.
Eq. (\ref{sm2}) allows us to express the 
excess entropy production, $\Delta s_\text{ex} = \Delta s_\text{m} - \Delta s_\text{hk}$
in terms of the last two terms in the second line.

Next, we note that Hatano and Sasa's \cite{HatanoSasa} derivation of their Eq. (11) 
can be easily adapted to our model active system resulting in the following relation
\begin{eqnarray}\label{HSgen}
\left<\exp\left[-\int_0^t dt' 
\dot{\lambda}(t')\partial_\lambda \phi(\mathbf{r},\varphi; \lambda)\right]\right>=1.
\end{eqnarray}

Combining Eqs. (\ref{sm2}) and (\ref{HSgen}) we get a version of the Hatano-Sasa 
relation,
\begin{eqnarray}\label{HS1} 
\left<\exp\left(-\Delta s_\text{ex} - \Delta \phi\right)\right>=1,
\end{eqnarray}
and then using the Jensen inequality we obtain a generalized Clausius relation,
\begin{eqnarray}\label{Cl1} 
-\left<\Delta s_\text{ex}\right> \ge \Delta \left<\phi\right>.
\end{eqnarray}
The right-hand-side of Eq. (\ref{Cl1}) can be rewritten as the change of the 
average (Shannon) entropy of the system,
\begin{eqnarray}\label{Dphi}
\Delta \left<\phi\right> = \Delta\left[-\int d\mathbf{r}d\varphi \,
p_s(\mathbf{r},\varphi;\lambda)\ln p_s(\mathbf{r},\varphi;\lambda)\right].
\end{eqnarray}
It can be shown that for a quasi-static process Eq. (\ref{Cl1}) becomes an equality. 

Recalling Eq.(\ref{sm2}) and the definition of $\Delta s_\text{hk}$ 
we can write the stochastic excess entropy production as   
\begin{eqnarray}\label{sex}
\Delta s_\text{ex}(t)= - \int_0^t dt' \left[
\partial_\mathbf{r} \phi(\mathbf{r},\varphi; \lambda)\cdot\dot{\mathbf{r}}(t')
+ \partial_\varphi \phi(\mathbf{r},\varphi; \lambda)\dot{\varphi}(t')
\right].
\nonumber \\
\end{eqnarray}
Here, the first term at the right-hand-side can be interpreted
as the excess entropy production in the medium. The second term, which has the form of
the excess entropy production in the reservoir coupled to the self-propulsions,
originates from non-trivial correlations between the particle's position and 
self-propulsion. 

Generalized Clausius inequality (\ref{Cl1}) extends the 2nd law of thermodynamics
to active matter. For its derivation it is essential to distinguish between
the total and the excess entropy production, since for quasistatic processes the
former is not well defined \cite{HatanoSasa}.

Physically, in the limit of very fast evolution of self-propulsions one expects
that an active system should be governed by some kind of effective thermodynamic 
description. In particular, one can show that in the limit of vanishing persistence
time at constant effective temperature an AOUP becomes equivalent to a Brownian 
particle. We note that in this limit, the joint steady-state distribution 
factorizes a product of the 
distributions of the position and of the self-propulsion, and, if the external
force is conservative, the former distribution acquires Gibbsian form. As a result,
generalized Clausius inequality (\ref{Cl1}) becomes the standard 
Clausius inequality. 

\textit{Integral fluctuation theorem for $\Delta s_\text{tot}$. --}
To derive the above described framework we did not
make any explicit assumptions regarding the behavior of the self-propulsion
under the time-reversal symmetry \cite{commenttrs}. We note that the identification
(\ref{sm1}) can be derived from the assumption that the self-propulsion is even 
under the reversal of time \cite{Shankar}. If we do use this assumption, we can 
derive an integral fluctuation theorem for the total entropy production
following Seifert's \cite{Seifert2005} derivation of the analogous theorem 
in standard stochastic thermodynamics. 

We consider a time dependent 
control parameter, \textit{i.e.} a protocol, $\lambda(t')$, and a reversed
protocol $\tilde{\lambda}(t')\equiv\lambda(t-t')$, and forward and reversed
trajectories, $\mathbf{x}(t')$ and $\tilde{\mathbf{x}}(t')\equiv\mathbf{x}(t-t')$,
where $\mathbf{x}\equiv [\mathbf{r},\varphi]$. It can be shown that the 
ratio of the probabilities of the forward and backward trajectories, conditioned on
their respective initial values, gives the stochastic entropy production
in the medium, Eq. (\ref{sm1}),
\begin{eqnarray}\label{ratio1}
\ln \frac{p[\mathbf{x}(t')|\mathbf{x}(0)]}
{\tilde{p}[\tilde{\mathbf{x}}(t')|\tilde{\mathbf{x}}(0)]} = \Delta s_\text{m}(t).
\end{eqnarray}
Next, we combine the left-hand-side of Eq. (\ref{ratio1}) with normalized
distributions of the initial values for the forward and reversed trajectories,
$p(\mathbf{x}(0))$ and $p(\tilde{\mathbf{x}}(0))$,
and we note that the latter distribution is equal to the distribution of the 
final values of the forward trajectory, 
$p(\tilde{\mathbf{x}}(0))=p(\mathbf{x}(t))$. 
In this way we get
\begin{eqnarray}\label{ratio2}
\ln \frac{p[\mathbf{x}(t')|\mathbf{x}(0)]p(\mathbf{x}(0))}
{\tilde{p}[\tilde{\mathbf{x}}(t')|\tilde{\mathbf{x}}(0)]p(\tilde{\mathbf{x}}(0))} 
= \Delta s_\text{tot}(t),
\end{eqnarray}
which leads to the integral fluctuation theorem for $\Delta s_\text{tot}$,
\begin{eqnarray}\label{stotift}
\left<\exp(-\Delta s_\text{tot})\right> &=& 
\sum_{\mathbf{x}(t'),\mathbf{x}(0)} \!\!\!
p[\mathbf{x}(t')|\mathbf{x}(0)]p(\mathbf{x}(0)) e^{-\Delta s_\text{tot}}
\nonumber \\ &=& 
\sum_{\tilde{\mathbf{x}}(t'),\tilde{\mathbf{x}}(0)} \!\!\!
p[\tilde{\mathbf{x}}(t')|\tilde{\mathbf{x}}(0)]p(\tilde{\mathbf{x}}(0))=1.
\end{eqnarray}

\textit{Perspective. --} 
We proposed a generalization of the standard stochastic thermodynamics framework
to active matter. We showed that many previously derived results, 
\textit{e.g.} stochastic total and housekeeping entropies, their fluctuation theorems, 
Harada-Sasa and Hatano-Sasa relations, appear naturally in the new framework.
This opens the way to experimental and computational studies of the 
thermodynamics of small active matter systems.
  
In our opinion, further work is needed in two directions. 
First, the framework presented here is 
naturally well adapted to describe systems in which both thermal noise and 
self-propulsion are important. Indeed, the housekeeping entropy production quantifies 
the non-equilibrium character of an active system, compared to the corresponding 
equilibrium system without any activity. 
On the other hand, athermal systems with rapidly 
varying self-propulsion (\textit{e.g.} an AOUP in the limit of vanishing persistence time)
are \emph{equivalent} to equilibrium systems. 
It would be interesting to develop a criterion quantifying how close a given
active matter system is to the equivalent equilibrium system. 
Second, it would be 
interesting to analyze the work done by both external and self-propulsion forces, and
to investigate whether a free energy-like quantity can be defined. We note that 
if an active system is close to the corresponding equilibrium system without any 
activity, one would try to follow Hatano and Sasa \cite{HatanoSasa} and 
use the temperature of the medium to define the free energy.
On the other hand, for an active system with rapidly varying self-propulsion, which
is close to an effective equilibrium system, one should probably use an effective
temperature to define the free energy. It is at present unclear how to interpolate
between these two cases.

\begin{acknowledgments}
I thank Elijah Flenner 
for comments on the manuscript. I gratefully acknowledge the support
of NSF Grant No.~CHE 1800282.
\end{acknowledgments}


\begin{thebibliography}{99}
\bibitem{friction} For a real self-propelled colloidal particle $\gamma_t$ and
$\gamma_r$ are related through a formula involving the radius of the particle 
but here we treat them as independent model parameters.
\bibitem{phistochentr} One should note that $\phi(t)$ is different from the
stochastic entropy $s(t)$; the former is defined in terms of steady-state 
distribution $p_s$ calculated for fixed instantaneous value of the control parameter,
$\lambda(t)$ whereas the latter is defined in terms of the solution of the 
Fokker-Planck equation for a given time dependence of the control parameter.
\end{thebibliography}
\end{document}